\documentclass[twoside,twocolumn,aps,sort,amsfonts,amssymb,amsmath,prb,floatfix,showpacs]{revtex4}
\usepackage[T1]{fontenc}
\usepackage[latin1]{inputenc}
\usepackage{graphicx}

\makeatletter

\providecommand{\tabularnewline}{\\}

\usepackage{dcolumn}
\usepackage{bm}

\makeatother
\begin{document}

\title{Weak inter-band coupling in $\textrm{Mg}^{10}\textrm{B}_{2}$: a
specific heat analysis}

\author{A. W\"alte$^{1}$}

\email{waelte@ifw-dresden.de}

\author{S.-L. Drechsler$^{1}$}

\author{G. Fuchs$^{1}$}

\author{K.-H. M\"uller$^{1}$}

\author{K. Nenkov$^{1,2}$}

\author{D. Hinz$^{1}$}

\author{L. Schultz$^{1}$}

\affiliation{$^{1}$IFW Dresden, Institute for Metallic Materials, P.O. Box 270116,
D-01171 Dresden, Germany\\
$^{2}$International Laboratory for High Magnetic fields and Low Temperatures,
95 Gajowicka Str., 53 421 Wroclaw, Poland}

\date{\today}


\begin{abstract}
The superconducting state of $\textrm{Mg}^{10}\textrm{B}_{2}$ is
investigated by specific heat measurements in detail. The specific
heat in the normal state is analyzed using a recently developed computer
code. This allows for an extraction of the electronic specific heat
in the superconducting state with high accuracy and a fair determination
of the main lattice features. One of the two investigated samples
shows a hump in the specific heat at low temperatures within the superconducting
state, accompanied by an unusual low value of the small gap, $\Delta_{\pi}\left(0\right)=1.32\textrm{ meV}$,
pointing to a very weak inter-band coupling. This sample allows for
a detailed analysis of the contribution from the $\pi$-band to the
electronic specific heat in the superconducting state. Therefore the
usual analysis method is modified, to include the individual conservation
of entropy of both bands. From analyzing the deviation function $D\left(t\right)$
of $\textrm{MgB}_{2}$, the theoretically predicted weak inter-band
coupling scenario is confirmed.
\end{abstract}
\maketitle

\section{Introduction}

The discovery\cite{nagamatsu01} of the unexpected high $T_{\textrm{c}}\approx39\textrm{ K}$
of $\textrm{MgB}_{2}$ was surprising for such a simple binary system
and has motivated much experimental and theoretical work in order
to understand the physics behind this superconductor. It turned out
that the electronic structure exhibits two bands crossing the Fermi
level. However from the relatively small electronic density of states
$N\left(0\right)\approx0.71\textrm{ states}/\left(\textrm{eV}\cdot\textrm{unit-cell}\right)$
at the Fermi level,\cite{an01,kong01,liu01,kortus01,choi02b} one
would not expect a notable $T_{\textrm{c}}$. Nowadays it is generally
accepted that the high energy vibrations here play the crucial role.\cite{bohnen01,yildirim01,osborn01,shukla03,heid03b,yanson04}
A closer examination of the electronic structure revealed an almost
decoupled state of the two prominent bands, with the $\pi$-band contributing
$\approx57$ \% and the $\sigma$-band $\approx43$ \% to the total
density of states.\cite{choi02,mazin03} Considerable effort was done
to quantify the role of these two bands for the superconductivity
from theory and experiment (for a review see Ref. \onlinecite{yanson04}).
The specific heat of the superconducting state is usually analyzed
in terms of a linear combination of two $\alpha$-models, with one
energy gap above and one below the BCS limit. Recently \citeauthor{dolgov05}
compared this approach to a two-band Eliashberg approach and found
it adequate particularly for the case of enhanced inter-band scattering.\cite{dolgov05}
The effect of different inter-band scattering rates was analyzed by
\citeauthor{nicol05}, likewise based on two-band Eliashberg calculations.
These authors predict the visibility of a jump at $\approx0.2T/T_{\textrm{c}}$,
where the order parameter of the $\pi$-band is expected to change
strongest.\cite{nicol05}

In the present work we present the specific heat analysis of a $\textrm{Mg}^{10}\textrm{B}_{2}$-sample,
showing a small upturn in the same temperature region, indicating
a reduced inter-band coupling compared to so far published data. A
recently developed computer code for the analysis of lattice dynamics
from specific heat measurements in the normal state is used in order
to extract the specific heat in the superconducting state with high
accuracy. With respect to the relatively low inter-band coupling within
$\textrm{MgB}_{2}$ a slightly different approach to the superconducting
state is suggested. It involves one $\alpha$-model to describe the
jump height of the specific heat at $T_{\textrm{c}}$ using the condition
of entropy-conservation between $0<T<T_{\textrm{c}}$. Consequently
the remaining electronic specific heat in the superconducting state
can be ascribed to the phase-transition of the second band.

\section{Experimental}

Polycrystalline samples of $\textrm{Mg}^{10}\textrm{B}_{2}$ have
been prepared by solid-state reaction and a subsequent hot-pressure
treatment. To prepare the sample, a mixture of $\textrm{Mg}$ and
$^{10}\textrm{B}$ powder was pressed into a pellet, wrapped in $\textrm{Ta}$
foil and sealed in a quartz ampoule containing an $\textrm{Ar}$ atmosphere
at $180\textrm{ mbar}$. The sample was sintered for $2$ hours at
$950\textrm{ }^{\circ}\textrm{C}$. In order to obtain samples with
high $T_{\textrm{c}}$-values additional $10$ \% $\textrm{B}$ was
used. To counter the high volatility of $\textrm{Mg}$, an additional
small piece of pure $\textrm{Mg}$ was placed in the ampoule. The
hot-pressure treatment is necessary to reduce the porousness of the
derived sample. The obtained dense samples were characterized by x-ray
diffractometry to estimate their quality. No second phase peaks occurred,
indicating that even the contribution of the usually forming $\textrm{MgO}$
is negligible. The superconducting transition temperature (onset)
$T_{\textrm{c}}=40.2\textrm{ K}$ measured by ac-susceptibility is
in agreement with so far published values for $\textrm{Mg}^{10}\textrm{B}_{2}$.\cite{budko01,finnemore01}
The specific heat was measured in the temperature range of $0.3<T<200\textrm{ K}$
and in magnetic fields up to $\mu_{0}H=9\textrm{ T}$ using a Quantum
Design Physical Property Measurement System. In the following analysis
we denote the sample showing the low-temperature upturn with {}``A''
and the piece not showing this anomaly with {}``B''.

\section{Results and Analysis}

\begin{figure}
\begin{center}\includegraphics[%
  width=0.45\textwidth,
  keepaspectratio]{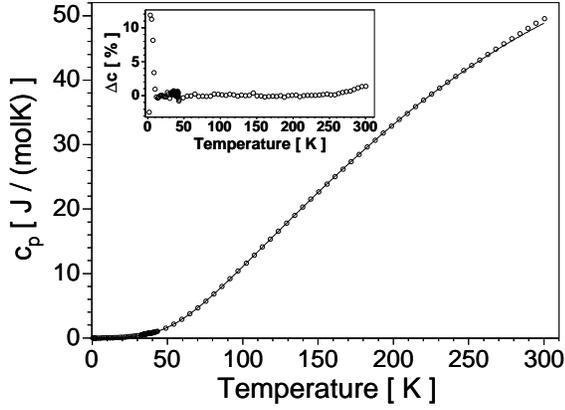}\end{center}

\caption{Specific heat of $\textrm{Mg}^{10}\textrm{B}_{2}$. Solid line: model
fit (see text for details). The superconducting transition is too
small to be visible in this plot. Inset: relative difference between
data and model (zero field data above $T_{\textrm{c}}$ and $\mu_{0}H=9\textrm{ T}$-data
below $T_{\textrm{c}}$). The large deviation of $\approx10$ \% at
low temperatures can be ascribed to the superconducting transition
which is not fully suppressed at $\mu_{0}H=9\textrm{ T}$.\label{fig:cp-300}}
\end{figure}
The specific heat of $\textrm{MgB}_{2}$ was analyzed in several experimental
works before. One crucial point in these analyses is the determination
of the Sommerfeld parameter, which is usually extracted by measuring
the specific heat in magnetic fields, which destroy the superconducting
state. However, analyses based on this approach can result in difficulties
concerning the conservation of entropy of the superconducting state,\cite{fisher03}
which is also visible in other quantities like the deviation function.\cite{nicol05}
It is unclear, if this is due to the experimental error,\cite{fisher03}
remnants of the superconducting signature of the field-measurements\cite{frederick01}
or perhaps due to a field-dependent Schottky anomaly, which is reported
for some samples including the present one.

In order to minimize such uncertainties we decided to apply a new
computer code to analyze the normal state of the specific heat in
a larger temperature region. The main ideas of the code have been
sketched in a previous work.\cite{waeltec} A more detailed description
will be published elsewhere. In summary our computer code makes use
of a linear combination of the well known Einstein and Debye models
for the specific heat in the normal state. The usual procedure in
fitting specific heat data using this code is to vary the Sommerfeld
parameter until the fit quality is optimized.

\begin{figure}
\begin{center}\includegraphics[%
  width=0.43\textwidth,
  keepaspectratio]{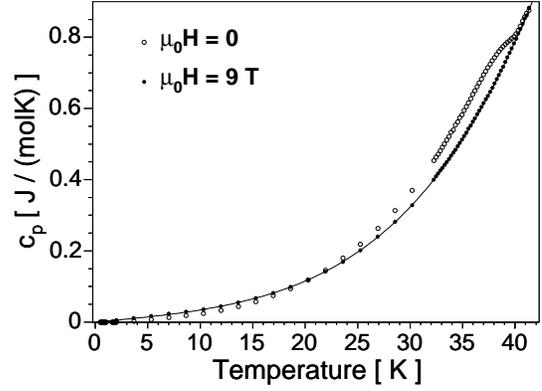}\end{center}

\caption{Specific heat of $\textrm{Mg}^{10}\textrm{B}_{2}$ at low temperatures.
Solid line: Model of the normal state specific heat (see text for
details).\label{fig:cp-40}}
\end{figure}

\subsection{Specific heat in the normal state}

The specific heat of the investigated $\textrm{Mg}^{10}\textrm{B}_{2}$
sample \textbf{A} is shown in Fig. \ref{fig:cp-300}. To extract the
electronic specific heat in the superconducting state, one has to
determine the Sommerfeld parameter and the lattice part. For this
purpose the zero-field data and $\mu_{0}H=9\textrm{ T}$ data measured
down to $T=20\textrm{ K}$ were used. The application of the mentioned
computer code indicated additional anharmonic effects starting at
$T\approx250\textrm{ K}$. Therefore the fitting temperature range
of the computer code was restricted to $20<T<250\textrm{ K}$. To
increase the accuracy of the fit, the conservation of entropy of the
superconducting transition was used as an additional requirement for
a successful fit. The Sommerfeld parameter was varied using a step-size
of $0.005\textrm{ mJ}/\left(\textrm{molK}^{2}\right)$ until the deviation
between the data and the fitted curve was minimized and the entropy
of the superconducting transition was conserved. At temperatures below
$\approx4\textrm{ K}$ one has also to account for hyperfine structure
contributions,\cite{bouquet01,yang01,wang01} given by a Schottky
model\[
c_{\textrm{s}}\left(T\right)=x\frac{Z\sum_{i}L_{i}^{2}\exp\left(L_{i}/T\right)-\left[\sum_{i}L_{i}\exp\left(L_{i}/T\right)\right]^{2}}{\left(TZ\right)^{2}},\]
with partition function $Z=\sum_{i}\left[\exp\left(L_{i}/T\right)\right]^{-1}$,
eigenvalues $L_{i}=\Delta_{i}/k_{\textrm{B}}$ and concentration $x$
of paramagnetic particles. The specific heat model is than given by
$c_{\textrm{p}}\left(T\right)=\gamma_{\textrm{N}}T+c_{\textrm{lattice}}\left(T\right)+c_{\textrm{s}}\left(T\right)$.
The result of the fitting procedure is shown as black line in Fig.
\ref{fig:cp-300}. The inset shows the difference between the data
and the model, which is well below $1$ \% except for low temperatures,
indicating that there the superconducting signature is not fully suppressed
by the applied $9\textrm{ T}$-field. Piece B shows nearly the same
data-to-model-difference which is not shown here.

Fig. \ref{fig:cp-40} shows the specific heat of sample A at $H=0$
and $\mu_{0}H=9\textrm{ T}$ in the low temperature region. The very
good agreement between the model and the field measurement is visible.
The Sommerfeld parameter converged to $\gamma_{\textrm{N}}=2.69\left(1\pm0.004\right)\textrm{ mJ}/\left(\textrm{molK}^{2}\right)$,
agreeing well with published values of high quality samples, ranging
from $\gamma_{\textrm{N}}\approx2.5-2.7\textrm{ mJ}/\left(\textrm{molK}^{2}\right)$.\cite{bouquet01,yang01,wang01,fisher03} 

Using the bare electron parameter $\gamma_{0}=\pi^{2}k_{\textrm{B}}^{2}N\left(0\right)/3=1.67\textrm{ mJ}/\left(\textrm{molK}^{2}\right)$,
the mean electron-phonon coupling constant can be calculated from
the mass enhancement relation\begin{equation}
\gamma_{\textrm{N}}=\gamma_{0}\left(1+\bar{\lambda}_{\textrm{ph}}\right)\label{eq:mass-enhancement}\end{equation}
as $\bar{\lambda}_{\textrm{ph}}\approx0.61$. The mean electron-phonon
coupling constant is related to the partial coupling constants in
the $\sigma$- and $\pi$-band by\begin{equation}
\bar{\lambda}_{\textrm{ph}}=\lambda_{\textrm{ph},\sigma}\frac{\gamma_{0,\sigma}}{\gamma_{0}}+\lambda_{\textrm{ph},\pi}\frac{\gamma_{0,\pi}}{\gamma_{0}}.\label{eq:lambda-mean}\end{equation}
From fitting the Schottky contribution $c_{\textrm{s}}\left(T\right)$
a concentration of $x=4.25\times10^{-4}\textrm{ mol}$ of paramagnetic
particles with two energy levels, $\Delta_{1}\approx0.014\textrm{ meV}$
and $\Delta_{2}\approx0.453\textrm{ meV}$ is estimated, probably
due to small amounts of $\textrm{Fe}$-impurities.\cite{wang01,yang01}
However a quantitative analysis of the energy levels requires measurements
below $T\approx0.3\textrm{ K}$, where the normal state contributions
vanish.

\begin{table}

\caption{Phonon-related parameters as obtained from the normal state specific
heat analysis. Numbering according to Fig. \ref{fig:pdos}.\label{tab:lattice-parameters}}

\begin{ruledtabular}

\begin{tabular}{ccccccccc}
&
&
\multicolumn{2}{c}{
acoustic --
}&
\multicolumn{5}{c}{
optical modes
}\tabularnewline
&
&

D1
&

D2
&

E1
&

E2
&

E3
&

E4
&

E5
\tabularnewline
\hline

$\omega$
&

$\left[\textrm{meV}\right]$
&

50.8
&

78.9
&

15.9
&

29.4
&

42.2
&

84.2
&

86.7
\tabularnewline

$\nu_{i}$
&
&

2
&

1
&

0.05
&

1
&

1
&

2
&

1.95
\tabularnewline
\end{tabular}

\end{ruledtabular}
\end{table}

\begin{figure}
\begin{center}\includegraphics[%
  width=0.45\textwidth,
  keepaspectratio]{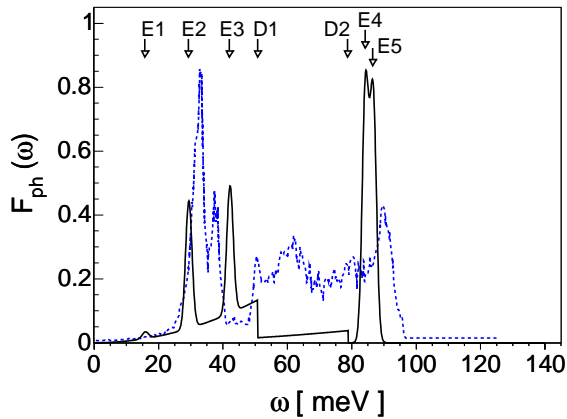}\end{center}

\caption{(Color online) Phonon density of states. Black line: result from
the specific heat analysis. Dashed line: calculated PDOS taken from
Ref. \onlinecite{yildirim01} (scaled to the peak value of the present
result). Numberings according to Tab. \ref{tab:lattice-parameters}.\label{fig:pdos}}
\end{figure}

The derived phonon-related parameters are summarized in Tab. \ref{tab:lattice-parameters}.
The resulting phonon density of states (PDOS) of sample A is shown
in Fig. \ref{fig:pdos} in comparison with a calculated PDOS taken
from Ref. \onlinecite{yildirim01}. The nice agreement, in particular
of the prominent modes at $\approx30\textrm{ meV}$ and $\approx85\textrm{ meV}$
($E_{2\textrm{g}}$) shows, that our computer code is able to extract
the main features of the phonon density of states from specific heat
measurements independently. A similar agreement is found from comparing
the specific heat derived PDOS with the generalized density of states
measured by neutron scattering experiments.\cite{osborn01,yildirim01,heid03b}

\subsection{Specific heat in the superconducting state}

\begin{figure}
\begin{center}\includegraphics[%
  width=0.42\textwidth,
  keepaspectratio]{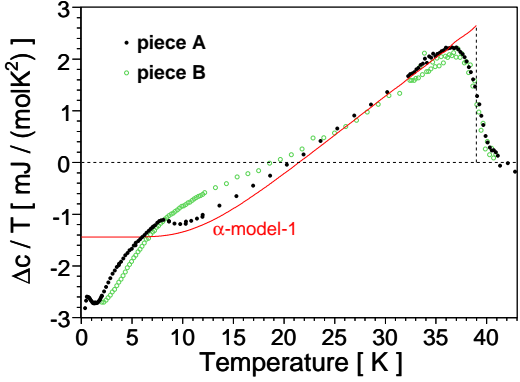}\end{center}

\caption{(Color online) Electronic specific heat $\Delta c/T$ of $\textrm{Mg}^{10}\textrm{B}_{2}$
in the superconducting state. Symbols: data of piece A. Broad line:
data of piece B. Solid line marked with {}``$\alpha$-model-1'':
$\alpha$-model fitted to the jump of piece A satisfying the condition
of entropy-conservation.\label{fig:jump}}
\end{figure}

Fig. \ref{fig:jump} shows the specific heat below $T=43\textrm{ K}$
after subtraction of the normal state contribution for the two investigated
samples A and B. The idealized specific heat jump of $\Delta c=103.35\textrm{ mJ}/\left(\textrm{molK}\right)$
for sample A is in the range of reported values, ranging from $\Delta c=81$
(Ref. \onlinecite{wang01}) to $133\textrm{ mJ}/\left(\textrm{molK}\right)$.\cite{bouquet01}
The local maximum at $T\approx9\textrm{ K}$ found for piece A has
not been reported before. The measurement was repeated on that piece
and the small hump is found to be reproduceable. Piece B from the
same initial sample does not show this hump, but a much more broadened
feature (open circles). Most probably the observed hump is related
to a relatively weak coupling between the two bands (a more detailed
analysis including magnetic field dependence is presented in section
\ref{sub:upturn-analysis}).\cite{nicol05,shulga_note} In view of
the two-band properties of $\textrm{MgB}_{2}$ it is more illustrative
to continue the analysis of piece A.

Assuming a very weak inter-band coupling, the jump at $T_{\textrm{c}}$
is nearly completely due to the $\sigma$-band which is responsible
for the large $T_{\textrm{c}}$. Consequently one $\alpha$-model
was used to fit the jump at $T_{\textrm{c},\sigma}=39.0\textrm{ K}$
using an entropy-conserving condition. This unambiguously results
in a Sommerfeld parameter $\gamma_{\textrm{N},\sigma}\approx1.44\textrm{ mJ}/\left(\textrm{molK}^{2}\right)$
of the $\sigma$-band, avoiding any uncertainty in the correct partitioning
of the Sommerfeld parameter under the two bands (see Ref. \onlinecite{bouquet01b}
for a short summary). The gap ratio amounts $2\Delta_{\sigma}\left(0\right)/\left(k_{\textrm{B}}T_{\textrm{c},\sigma}\right)=3.98$,
resulting in $\Delta_{\sigma}\left(0\right)=6.70\textrm{ meV}$.%

\begin{figure}
\begin{center}\includegraphics[%
  width=0.42\textwidth,
  keepaspectratio]{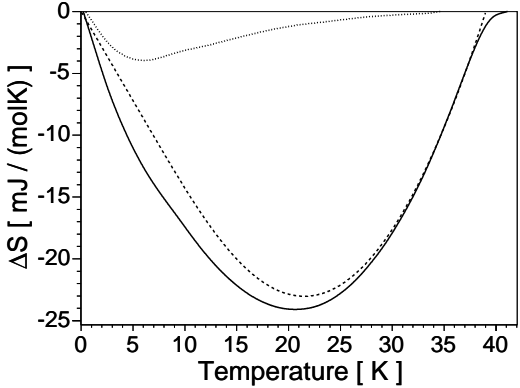}\end{center}

\caption{Entropy difference between the normal and superconducting state of
sample A. Solid line: entropy difference determined from the electronic
specific heat. Dashed line: entropy difference of the fitted $\alpha$-model.
Dotted line: residual entropy difference.\label{fig:entropy}}
\end{figure}

Fig. \ref{fig:entropy} shows the entropy difference of the electrons
between the normal and the superconducting state. The solid line represents
the entropy difference of the experimental data, agreeing well with
results obtained by \citeauthor{yang01}\cite{yang01} The dashed
line is the entropy difference of the fitted $\alpha$-model and the
dotted line corresponds to the residual entropy difference which can
be attributed to the $\pi$-band. The superconducting transition of
the $\pi$-electrons is strongly smeared due to the inter-band coupling,
but the main part of the $\pi$-electrons seems to become superconducting
at $T_{\textrm{c}}^{^{\star}}\approx9\textrm{ K}$, otherwise the
minimum in the entropy difference (dotted line) would be expected
at higher temperatures. This interpretation does not mean that the
gap of the $\pi$-band also opens at $T_{\textrm{c}}^{\star}$. In
fact, any strength of inter-band coupling results in one single critical
temperature $T_{\textrm{c},\sigma}$ for both gaps. However the order
parameter of the $\pi$-band is expected to show its strongest change
around $T_{\textrm{c}}^{\star}$.\cite{nicol05}

\begin{figure}
\begin{center}\includegraphics[%
  width=0.42\textwidth,
  keepaspectratio]{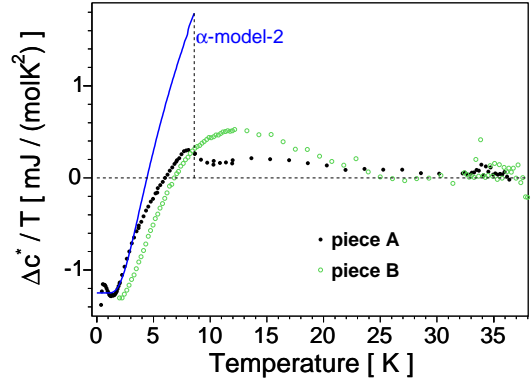}\end{center}

\caption{(Color online) Electronic specific heat in the superconducting state
which can be attributed to the $\pi$-electrons. The normal state
and the contribution of the $\sigma$-band described by the $\alpha$-model
fitted to the jump have been subtracted. Filled circles: sample A.
Open circles: sample B. Solid line marked by {}``$\alpha$-model-2'':
$\alpha$-model fitted to the low-temperature part of the small hump
of piece A satisfying the condition of entropy-conservation.\label{fig:small-jump}}
\end{figure}

Fig. \ref{fig:small-jump} shows the specific heat $\Delta c^{\star}/T$
in the superconducting state of sample A (filled circles) after subtraction
of the $\sigma$-band contribution given by the $\alpha$-model (see
Fig. \ref{fig:jump}). The Sommerfeld parameter of the $\pi$-band
amounts $\gamma_{\textrm{N},\pi}=\gamma_{\textrm{N}}-\gamma_{\textrm{N},\sigma}=1.25\textrm{ mJ}/\left(\textrm{molK}^{2}\right)$.
The small deviations for $T<2\textrm{ K}$ can be attributed to the
inexact knowledge of the Schottky contribution, which should be determined
at much lower temperatures. However the accuracy is sufficient for
the present analysis. A second $\alpha$-model was applied to the
remaining specific heat, using $T_{\textrm{c}}^{\star}\approx9\textrm{ K}$.
Since the superconducting transition of the $\pi$-electrons is strongly
smeared (up to $T_{\textrm{c},\sigma}$), reliable results on the
energy gap can be derived only from the low temperature tail. Using
$\gamma_{\textrm{N}}=\gamma_{\textrm{N},\pi}$ and $T_{\textrm{c}}^{\star}=8.7\textrm{ K}$
the solid line shown in Fig. \ref{fig:small-jump} (marked by {}``$\alpha$-model-2'')
is derived. The resulting gap amounts $\Delta_{\pi}\left(0\right)=1.32\textrm{ meV}$
with the BCS gap ratio $2\Delta_{\pi}\left(0\right)/\left(k_{\textrm{B}}T_{\textrm{c}}^{\star}\right)=3.52$.
This value is much smaller than reported from other specific heat
analyses, which gave $\Delta_{\pi}\left(0\right)=1.7-2.2\textrm{ meV}$.\cite{bouquet01,wang01,fisher03}
It is interesting to compare the small gap of sample A with that of
sample B. The open circles in Fig. \ref{fig:small-jump} show the
remaining specific heat of sample B after subtracting an $\alpha$-model,
fitted in a similar way to the jump height (resulting in $\Delta_{\sigma}\left(0\right)=6.38\textrm{ meV}$
for the larger gap). A fit of another $\alpha$-model to the low-temperature
part of $\Delta c^{\star}/T$ (not shown here) as it was done for
sample A, results in $\Delta_{\pi}\left(0\right)=1.76\textrm{ meV}$
for the small gap. A successful fit also requires a larger $T_{\textrm{c}}^{\star}=11.5\textrm{ K}$
for this sample, comparable to $T_{\textrm{c}}^{\star}\approx11\textrm{ K}$
reported by \citeauthor{bouquet01}\cite{bouquet01} The smaller gap
found for the $\pi$-band of sample A supports the scenario of a reduced
inter-band coupling in this sample.\cite{nicol05} The gap parameters
are summarized in Tab. \ref{tab:gap-parameter}.

\begin{table}

\caption{Gap values obtained from the $\alpha$-model for the two investigated
samples.\label{tab:gap-parameter}}

\begin{center}\begin{ruledtabular}\begin{tabular}{ccccc}
Sample&
$\Delta_{\sigma}\left[\textrm{meV}\right]$&
$\Delta_{\pi}\left[\textrm{meV}\right]$&
$T_{\textrm{c},\sigma}\left[\textrm{K}\right]$&
$T_{\textrm{c}}^{\star}\left[\textrm{K}\right]$\tabularnewline
\hline
A&
6.70&
1.32&
38.9&
8.7\tabularnewline
B&
6.38&
1.76&
38.9&
11.5\tabularnewline
\end{tabular}\end{ruledtabular}\end{center}
\end{table}

\begin{figure}
\begin{center}\includegraphics[%
  width=0.48\textwidth,
  keepaspectratio]{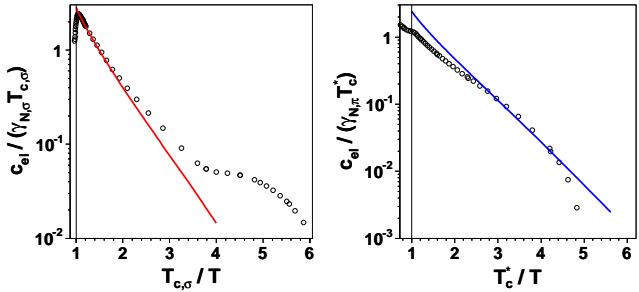}\end{center}

\caption{(Color online) Normalized electronic specific heat of sample A. Left
panel: experimental electronic specific heat of the superconducting
state. Solid line: $\alpha$-model according to Fig. \ref{fig:jump}.
Right panel: remaining electronic specific heat attributed to the
$\pi$-band, normalized on $T_{\textrm{c}}^{\star}=8.7\textrm{ K}$.
Solid line: $\alpha$-model according to Fig. \ref{fig:small-jump}.
\label{fig:bcs}}
\end{figure}

In Fig. \ref{fig:bcs}, the electronic specific heat of the $\sigma$-
and the $\pi$-band in the superconducting state are compared. The
left panel shows the specific heat data normalized on $T_{\textrm{c},\sigma}$
and $\gamma_{\textrm{N},\sigma}$ of the $\sigma$-band. The right
panel shows the remaining electronic specific heat as given in Fig.
\ref{fig:small-jump} (plus $\gamma_{\textrm{N},\pi}$) normalized
on $T_{\textrm{c}}^{\star}$ and $\gamma_{\textrm{N},\pi}$ of the
$\pi$-band. The deviation for $T_{\textrm{c}}^{\star}/T>4.5$ between
the data and the corresponding $\alpha$-model is due the mentioned
uncertainty of the Schottky contribution (see also Fig. \ref{fig:small-jump}).

\subsection{Field-dependence of the specific heat below $15\textrm{ K}$\label{sub:upturn-analysis}}

\begin{figure}
\begin{center}\includegraphics[%
  width=0.45\textwidth,
  keepaspectratio]{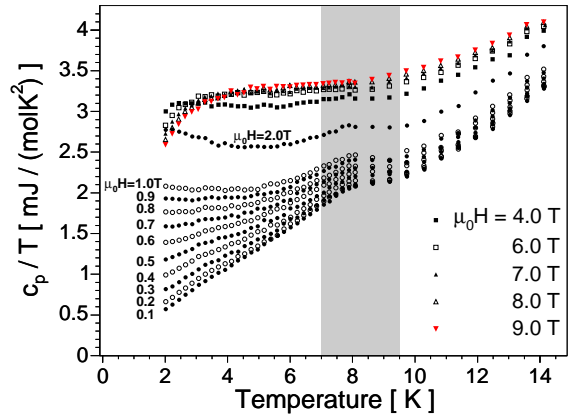}\end{center}

\caption{(Color online) Specific heat $c_{\textrm{p}}/T$ in different applied
magnetic fields as labeled. The vicinity of the upturn observed in
the zero-field measurement is marked by the shaded area.\label{fig:cp-field}}
\end{figure}

In order to clarify the nature of the hump found in the electronic
specific heat in the superconducting state near $8\textrm{ K}$, specific
heat measurements in applied magnetic fields between $\mu_{0}H=0.1$
and $9.0\textrm{ T}$ have been performed. The resulting specific-heat-to-temperature-ratio
is shown in Fig. \ref{fig:cp-field}. The shaded area marks the temperature
region of the anomaly for the zero-field measurement as shown in Fig.
\ref{fig:jump}. Two properties of the anomaly can be derived from
these measurements. First, the position of the anomaly is more or
less unchanged. Second, the anomaly is flattened out with increasing
magnetic field, accompanied by an increase of the specific heat at
low temperatures. The anomaly vanishes at approximately $9\textrm{ T}$,
where the superconducting signal of the specific heat is very hard
to separate, indicating that the origin of the anomaly is connected
to the superconductivity of $\textrm{Mg}^{10}\textrm{B}_{2}$. Comparing
the hump qualitatively with recent calculations performed by \citeauthor{nicol05},\cite{nicol05}
it seems that the inter-band coupling in our sample is reduced by
$\approx30-50$ \% compared to previously reported samples (see section
\ref{sec:discussion} for a detailed analysis). The inter-band coupling
can therefore be expected to be in the order of $\approx0.2$. Comparing
the specific heat measurements of \citeauthor{bouquet01}\cite{bouquet01}
and \citeauthor{wang01},\cite{wang01} as was done in Ref. \onlinecite{bouquet01b}
it is obvious that it is possible to vary the inter-band coupling,
probably by means of the preparation technique. In particular the
low temperature anomaly in the data of \citeauthor{bouquet01} is
much more pronounced than in the data of \citeauthor{wang01}, but
there is no upturn visible as in our sample (A).

\section{Discussion\label{sec:discussion}}

Both gap values for sample A, $\Delta_{\sigma}\left(0\right)=6.70\textrm{ meV}$
and $\Delta_{\pi}\left(0\right)=1.32\textrm{ meV}$ are in the lower
region of the expected values for $\textrm{MgB}_{2}$.\cite{yanson04}
In general, specific heat measurements are expected to result in averaged
gap values, resulting from the two bands. On the other hand, transport
measurements like point-contact spectroscopy measurements are also
influenced by the Fermi velocity distribution of the charge carriers
of the individual bands, usually resulting in gap values from the
faster charge carriers of the two bands. A comparison with calculations
by \citeauthor{choi02}\cite{choi02} shows that the present gap values
represent the lower end of the calculated gap distribution.

Additional information on the electron-phonon coupling can be derived
by comparing the resulting thermodynamic quantities with band-structure
results. For this purpose we make use of a strong coupling correction
term given by \citeauthor{carbotte90}:\cite{carbotte90}\[
\frac{\Delta c}{\gamma_{\textrm{N}}T_{\textrm{c}}}=1.43\left(1+\frac{53}{x^{2}}\ln\frac{x}{3}\right)\]
with $x=\omega_{\textrm{ln}}/T_{\textrm{c}}$. Analyzing the jump
with $T_{\textrm{c},\sigma}$ and $\gamma_{\textrm{N},\sigma}$ in
this manner, the characteristic phonon frequency amounts to $\omega_{\textrm{ln},\sigma}=712\textrm{ K}$.
Using the McMillan formula (refined by \citeauthor{allen75}\cite{allen75}):\begin{equation}
T_{\textrm{c}}=\frac{\omega_{\textrm{ln}}}{1.2}\exp\left[-\frac{1+\lambda_{\textrm{ph}}}{\lambda_{\textrm{ph}}-\mu^{\star}\left(1+0.6\lambda_{\textrm{ph}}\right)}\right],\label{eq:Allen-Dynes}\end{equation}
with the Coulomb pseudopotential $\mu^{\star}=0.10$, one finds $\lambda_{\textrm{ph},\sigma}=0.82$.
With $\gamma_{\textrm{N},\sigma}=1.44\textrm{ mJ}/\left(\textrm{molK}^{2}\right)$
one gets $\gamma_{0,\sigma}=0.79\textrm{ mJ}/\left(\textrm{molK}^{2}\right)$
for the bare electron parameter of the $\sigma$-band. With $\gamma_{0}=1.67\textrm{ mJ}/\left(\textrm{molK}^{2}\right)$,
the bare electron parameter for the $\pi$-band amounts to $\gamma_{0,\pi}=\gamma_{0}-\gamma_{0,\sigma}=0.88\textrm{ mJ}/\left(\textrm{molK}^{2}\right)$.
The electron-phonon coupling within the $\pi$-band can then be estimated
from the mass enhancement (Eq. \ref{eq:mass-enhancement}) as $\lambda_{\textrm{ph},\pi}=\gamma_{\textrm{N},\pi}/\gamma_{0,\pi}-1=0.42$.
The corresponding characteristic phonon frequency determined from
Eq. (\ref{eq:Allen-Dynes}) amounts to $\omega_{\textrm{ln},\pi}=1270\textrm{ K}$.
It can only be guessed whether this rather large value is related
to the mentioned anharmonic effects seen in the specific heat at large
temperatures or inaccuracies of the band structure calculations. From
this analysis the $\sigma$-band contributes with $\approx47$ \%
and the $\pi$-band with $\approx53$ \% to the total density of states.
Setting the derived parameters in Eq. (\ref{eq:lambda-mean}), one
confirms the value of $\bar{\lambda}_{\textrm{ph}}=0.61$. The obtained
results are summarized in Tab. \ref{tab:e-p-parameter} in comparison
with calculated values derived by analyzing theoretical Eliashberg
functions.\cite{golubov02b}

\begin{figure}
\begin{center}\includegraphics[%
  width=0.50\textwidth,
  keepaspectratio]{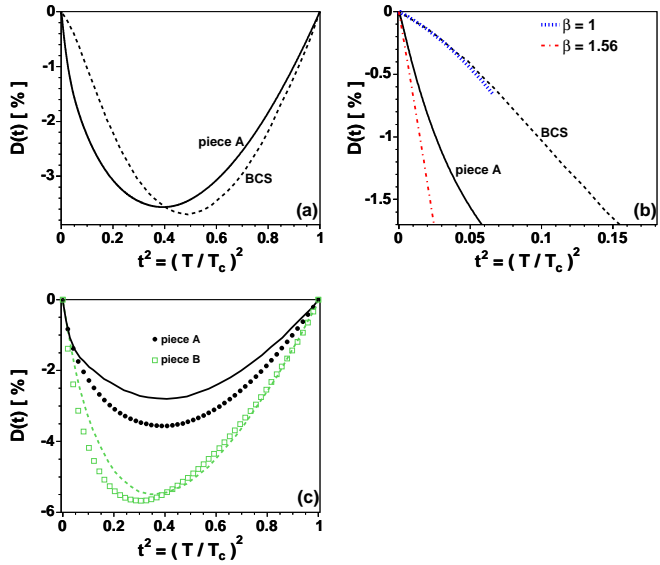}\end{center}

\caption{(Color online) Experimental deviation function $D\left(t\right)$
compared to model predictions. (a) and (b): $D\left(t\right)$ of
sample A compared to the BCS-model prediction. Dotted line in (b):
Eq. (\ref{eq:D(t)}) with $\beta=1$ (BCS case). Dash-dotted line
in (b): Eq. (\ref{eq:D(t)}) with $\beta$ recalculated according
to Eq. (\ref{eq:beta}), using the experimentally derived values.
(c): Comparison of $D\left(t\right)$ of both samples with Eliashberg-model
calculations taken from Ref. \onlinecite{nicol05}. Dashed line: calculation
using {}``normal'' inter-band coupling. Solid line: calculation
with $50$ \% reduced inter-band coupling.\label{fig:deviation}}
\end{figure}

A nice check of the obtained parameters can be performed by analyzing
the deviation function $D\left(t\right)$ given by (see for example
Ref. \onlinecite{poole1995}):\[
D\left(t\right)=\frac{H_{\textrm{c}}\left(T\right)}{H_{\textrm{c}}\left(0\right)}-\left(1-t^{2}\right),\]
with $t=T/T_{\textrm{c}}$ (in the following $T_{\textrm{c}}=T_{\textrm{c},\sigma}$
is used). It gives the deviation from the two-fluid model of the superconducting
state, thus revealing important information on the superconducting
state. Fig. \ref{fig:deviation}(a) shows that $D\left(t\right)$
of $\textrm{Mg}^{10}\textrm{B}_{2}$ (sample A) closely resembles
the BCS prediction, in particular near $T_{\textrm{c}}$. However,
there are some deviations from this behavior. First the minimum is
shifted to lower temperatures and second, the behavior for $T\rightarrow0$
strongly deviates from the BCS prediction.

\begin{table}

\caption{Parameters characterizing the electron-phonon coupling. The first
two columns contain calculated values\cite{golubov02b}. The characteristic
phonon frequencies have been calculated by integrating the corresponding
Eliashberg functions in Ref. \onlinecite{golubov02b}. $T_{\textrm{c}}$
was calculated using Eq. (\ref{eq:Allen-Dynes}) with $\mu^{\star}=0.1$
(thereby ignoring all inter-band effects). The next two columns show
the present experimental results for sample A.\label{tab:e-p-parameter}}

\begin{ruledtabular}\begin{tabular}{cccccc}
&
&
\multicolumn{2}{c}{
theoretical --
}&
\multicolumn{2}{c}{
experimental values
}\tabularnewline
&
&

$\pi$
&

$\sigma$
&

$\pi$
&

$\sigma$
\tabularnewline
\hline

$\omega_{\textrm{ln}}$
&

$\left[\textrm{K}\right]$
&

668
&

770
&

1270
&

712
\tabularnewline

$\lambda_{\textrm{ph}}$
&
&

0.45
&

1.02
&

0.42
&

0.82
\tabularnewline

$T_{\textrm{c}}$
&

$\left[\textrm{K}\right]$
&

6.25
&

59.33
&

8.7
&

38.9
\tabularnewline
\end{tabular}

\end{ruledtabular}

\end{table}
Based on the weak-coupling two-band equations derived by \citeauthor{moskalenko73}\cite{moskalenko73}
and by analyzing numerical data given in Ref. \onlinecite{carbotte90}
we derived an analytical expression for $D\left(t\right)$, describing
deviations due to two-band influences. Within the BCS theory the dependence
of the thermodynamic critical field for $T\rightarrow0$ is given
by\begin{equation}
\left[\frac{H_{\textrm{c}}\left(T\right)}{H_{\textrm{c}}\left(0\right)}\right]^{2}=1-2.12\beta\left(\frac{T}{T_{\textrm{c}}}\right)^{2},\label{eq:D(t)}\end{equation}
with $\beta\approx1$. In the case of increasing electron-phonon coupling
$\beta$ is decreasing but it is increasing again in the case of two-band
influence with different gaps in both bands. In this case $\beta$
can be expressed as\begin{equation}
\beta=z\frac{B\left(\frac{\omega_{\textrm{ln},\sigma}}{T_{\textrm{c}}}\right)+B\left(\frac{\omega_{\textrm{ln},\pi}}{T_{\textrm{c}}}\right)}{z^{2}+v}\exp\left[\frac{z^{2}-v}{z^{2}+v}\ln\left(z\right)\right],\label{eq:beta}\end{equation}
with\begin{eqnarray*}
z & = & \frac{\Delta_{\sigma}\left(0\right)}{\Delta_{\pi}\left(0\right)},\\
v & = & \frac{\left(1+\lambda_{\textrm{ph},\pi}\right)N_{\pi}\left(0\right)}{\left(1+\lambda_{\textrm{ph},\sigma}\right)N_{\sigma}\left(0\right)},\\
B\left(x\right) & = & 1-\frac{7.5}{x^{2}}\ln\frac{x}{3}.\end{eqnarray*}
 Eq. (\ref{eq:beta}) corresponds to the BCS limit for the weak-coupling
single-band case and reproduces the strong coupling corrections reported
by \citeauthor{carbotte90}.\cite{carbotte90} The agreement of Eq.
(\ref{eq:beta}) with the two-band Eliashberg theory was checked numerically
for the theoretical case given by \citeauthor{golubov02b},\cite{golubov02b}
where the deviation is found to be less than $4$ \%.\cite{shulga_note}
Parameters obtained from these calculations are given in Tab. \ref{tab:e-p-parameter}.
Using the above derived electron-phonon coupling quantities and $T_{\textrm{c}}=T_{\textrm{c},\sigma}$,
one derives $\beta=1.56$. The dependence of $D\left(t\right)$ at
low temperatures is plotted in Fig. \ref{fig:deviation}(b), showing
the very nice agreement of this analytical formula with the experimental
deviation function.%

However, Fig. \ref{fig:deviation}(c) shows that the deviation function
of the present sample A (filled circles) looks quite different from
that of sample B (open squares). For the data of \citeauthor{bouquet01}\cite{bouquet01}
a minimum value of $\approx5.5$ \% was determined,\cite{nicol05}
similar to sample B, but sample A has a minimum value of only $\approx3.5$
\%.

It is interesting to compare these data to Eliashberg-model calculations.
\citeauthor{nicol05} recently published such calculations where they
varied the inter-band coupling, using the standard set of $\lambda_{\sigma\pi}=0.213$
and $\lambda_{\pi\sigma}=0.155$ and a second set with halved inter-band
$\lambda$-values. Fig. \ref{fig:deviation}(c) shows the comparison
between these calculations and our data. From that it can be concluded,
that the inter-band coupling in sample A is reduced compared to that
of sample B by approximately $30$ \%.

\begin{figure}
\begin{center}\includegraphics[%
  width=0.48\textwidth,
  keepaspectratio]{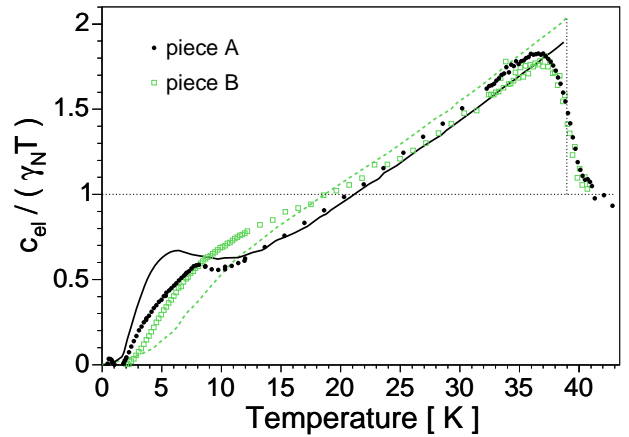}\end{center}

\caption{(Color online) Electronic specific heat $c_{\textrm{el}}/\gamma_{\textrm{N}}T$
of both samples compared to Eliashberg-calculations by \citeauthor{nicol05}.\cite{nicol05}
The dashed and solid lines correspond to the lines in Fig. \ref{fig:deviation}(c).\label{fig:eliashberg}}
\end{figure}

Fig. \ref{fig:eliashberg} shows a comparison between the electronic
specific heat below $T_{\textrm{c}}$ for the investigated two samples
and the calculations by \citeauthor{nicol05} {[}which correspond
to Fig. \ref{fig:deviation}(c){]}. There is a qualitative agreement
between the data sets and the calculations except for the vicinity
of $T_{\textrm{c}}$, where the calculations predict a reduction of
the normalized jump with decreasing inter-band coupling. In addition
there is no increase of $T_{\textrm{c}}$ for the sample with the
stronger inter-band coupling. However, \citeauthor{nicol05} showed
that for some combinations of $\lambda_{\pi\sigma}$ and $\lambda_{\sigma\pi}$,
even a decrease of $T_{c}$ is possible.\cite{nicol05} Therefore
it seems difficult to make predictions concerning $T_{\textrm{c}}$
or the jump.

The question, why two samples of the same initial piece should have
different values of the inter-band coupling constants is still unsolved.
In the present case it is most probably related to different amounts
of impurities, which influence the inter-band coupling by local lattice
distortions. These lattice distortions can modify the electron-phonon
coupling. Therefore it is also difficult to draw conclusions from
the Sommerfeld parameter, which is at the same time determined by
the two intra-band and inter-band coupling constants.

\section{Conclusion}

In the present work we report the first detailed specific heat analysis
of a $\textrm{MgB}_{2}$-sample with a particularly low inter-band
coupling. The observed low-temperature upturn within the superconducting
state is in accord with theoretical studies.\cite{nicol05} It is
shown, that just by using results of band-structure calculations,
meaningful physical quantities can be derived from specific heat measurements.
The relatively low gap values, $\Delta_{\sigma}\left(0\right)=6.38\textrm{ meV}$
and $\Delta_{\pi}\left(0\right)=1.76\textrm{ meV}$ of sample B, which
represents a {}``standard'' $\textrm{MgB}_{2}$-sample, naturally
emerge from the type of measurement. Transport measurements usually
result in higher values of the gaps, indicating that the gap values
depend on the velocity of the charge carriers, again raising the question,
if a gap distribution within the two bands as calculated by \citeauthor{choi02}
can become experimentally visible.\cite{choi02,choi02b,mazin04,choi04}
The resulting electron-phonon coupling constants for sample A show,
that the $\sigma$-band couples stronger than the $\pi$-band, as
expected, but still $\textrm{Mg}^{10}\textrm{B}_{2}$ can be considered
as a weak to medium coupling superconductor. Thus, the large characteristic
phonon frequency is the main reason for the relatively large superconducting
transition temperature, showing the crucial role of the $E_{2\textrm{g}}$-mode.
The obtained parameters for sample A were found to be in agreement
with the low temperature behavior of the deviation function $D\left(t\right)$.
Comparing the deviation function and the electronic specific heat
of the superconducting state of both samples with calculations by
\citeauthor{nicol05}, the scenario of a low inter-band coupling for
the present sample A is supported. This is in particular agreeing
well with the $\approx30$ \% smaller gap within the $\pi$-band of
that sample. This reduction was predicted theoretically in the case
of a reduced inter-band coupling constant by $\approx30$ \%.\cite{nicol05}
However sample B from the same initial piece does not show this peculiarity.
At present it is still unclear, which parameter in the sample preparation
process controls the inter-band coupling.

\begin{acknowledgments}
The DFG (SFB 463) is gratefully acknowledged for financial support.
We thank S. V. Shulga for fruitful discussions.
\end{acknowledgments}
\bibliographystyle{/work/exp/mgb2/paper/apsrev}
\bibliography{/work/exp/mgb2/paper/master}

\end{document}